\documentstyle[11pt]{article}
\begin{document}
\thispagestyle{empty}

\begin{center}

RUSSIAN GRAVITATIONAL ASSOCIATION\\
CENTER FOR SURFACE AND VACUUM RESEARCH\\
DEPARTMENT OF FUNDAMENTAL INTERACTIONS AND METROLOGY\\
\end{center}
\vskip 4ex
\begin{flushright}                              RGA-CSVR-016/94\\

                                               hep-ph@xxx.lanl.gov
\end{flushright} \vskip 45mm

\centerline {\large \bf        TEST  OF UNIVERSALITY HYPOTHESIS }
\centerline {\large \bf FOR SCALAR CONFINING POTENTIAL }
\centerline {\large  \bf BETWEEN QUARKS}
\vskip3mm \centerline   {       S.V.Semenov, V.V.Khruschev }
\centerline{            Surface and Vacuum Research Centre,}
\centerline             {Moscow, Russia}
\section{                   Introduction }
\noindent
\begin{abstract}
    The universality hypothesis for scalar confining potential
between quark and antiquark with different flavours is confirmed
within 0.05 level of relativity errors.
\end{abstract}
    Characteristic feature of quarks interaction potential is its
growth at high distances. This behaviour of potential requires the
clearification  at least the following questions: how long the inrease of the
potential takes place, what are its transformational properties according to
Lorentz group transformations, how the form of the increasing potential depends
on the quark flavours ?  In the present work  we shall consider in the
main the last question, supposing , and it is sufficiently well founded,
that one can believe the cofinement potential in some domain of
radius $\sim$ 1 Fm  to be linearly growing and scalar according to Lorentz
group transformations [1-3].  It is well known, that direct application of
    QCD for determination of spectra of bound states of quarks and/or
antiquarks leads to a number of difficulties, therefore the compound quark
models are frequently used, in which frameworks the sufficiently simple
calculations,in particular, of masses and decay widths can be produced (for
example,[1-3]).  The most simple, from the models, used for description of
    spectroscopic characteristics of hadrons, are the potential ones and
,especially, nonrelativistic potential models, which turned out to be
exceptionally succesive for the description of heavy quarkonia
characteristics. However, the presence of a number of nonperturbative
effects, connected, for instance, with the complicated structure of QCD
vacuum, and also, with the necessity to take into consideration the
nonrelativistic corrections, led to the complication of the formulation of
these models, and , as a rule, to the change to the worse of the precision of
obtained results. Undoubtedly, those models present the greatest interest,
which do not contradict the basic principles of QCD, have the minimal number
of phenomenological parameters, promote the clearing up of the interaction
form in nonperturbative domain and keep in calculation  of both light and
heavy hadrons characteristics the high level of precision, which have been
achieved in nonrelativistic models of heavy quarkonia.
    In the present work the phenomenological model of hadrons, proposed
in refs [4-5] is used, the basic statements of the model are given briefly
in section 2. As has been shown in section 3, the results of calculations
light mesons masses built of $u-, d-$quarks and antiquarks in the
framework of this model, in the approximation of zero current masses of
quarks and absence of static quasi-Coulomb interaction, are in good
agreement with experimental data. In the section 4 on the base of numerical
solution of the basic equation of the model with the account for the
quasi-Coulomb interaction the values of the strong interaction constants
in the perturbative domain in the presence of the quasi-Coulomb interaction
and in nonperturbative domain in the presence of scalar linearly growing
potential are found. In the last section, taking into account the
obtained results, the application of the universality hypothesis of the
scalar confining potential in the framework of the considered model
is discussed.

\section{          Phenomenological model of generalized
                     quark fields for hadrons}
\noindent

     In the framework of the model proposed in refs [4-5],  hadron is
the compound system  which consists of the general self-consistent confining
field, this field, likely, includes the production and annihilation effects
of sea $(q\overline{q})$- pairs and some number of valent quarks and/or
antiquarks.  For the sake of simplicity, in further computations we shall
suppose, that confining field is spherically symmetric and its space-time
motion is given by the centre position with coordinates $ q_{0}^{\mu}  ,\mu
=0,1,2,3$ The coordinates of valent quarks and/or antiquarks we shall write
as $q^\mu_i, \mu =0,1,2,3; i=1,\ldots,n$, where $n$ is the total number of
 valent fermion constituents.  We shall work in quasi-independent particles
 approximation, when the wave function of the whole system is represented as
    a product of its components wave functions, dependent in general on
coordinates of two or more constituents.  \begin {equation}
           \Psi_a(q_0,q_1,\ldots,q_n)=\Psi_0(q_0,q_1,\ldots,q_n)
\Psi_1(q_0,q_1)\ldots\Psi_n(q_0,q_n)
\end{equation}
\noindent
where wave function $ \Psi_0 (q_0,q_1,\ldots,q_n)$ describes the general field
state with the centre in point $q^\mu_0$, and $\Psi_i (q_0,q_i)$-is the state
of the
$i$-th valent fermion particle in this general field. It should be noticed,
that we consider the $i$-th particle not to be free, but bound with the
others by means of general field. Making the further simplifications, we
shall suppose, that the origin of the general field is due only to the
fermion components and combine the inertia centre of fermion components
system $Q^\mu$  , cannonically conjugate to their total momentum, with the
point
$q^\mu_0$     .
    We shall restrict the consideration of ordinary hadrons to the case [4],
when the behaviour of $\Psi_i (Q,q_i)$ in the confinement region is determined,
in
the main, by the Dirac equation with the scalar potential $U_i ((q_i -Q)^2 )$:
     .
\begin{equation}
(i \gamma^ \mu \frac{\partial}{\partial(q^\mu_i -Q^\mu )}-U_i
((q_i -Q)^2 )-\mu_i )\Psi_i (Q,q_i)=0
\end{equation}

    The dependence of $ \Psi_i(Q,q_i)$ on $q_i -Q$, is the only important one
for masses
spectra calculation. In the hadron rest-frame we shall use the well known
in many-particles theory single-time approximation,when $Q^0=q^0_1
=\ldots=q^0_1$
In that case the potentials $U_i ((q_i -Q)^2 )$ do not depend on time and it is
 possible to come to the standard Dirac equations solution for $i=1,\ldots,n$
    Thus, the hadron state in the given model is determined by the state of
the each quark, situated in its orbital, and by the state of confining
field, which interaction with the $i$-th quark is given by the potential
$U_i ((\vec{q_i}-\vec{Q_i}$.It should be noticed, that the general field is
assumed to be
"white",  unlike the colour gluon field, the quarks interact with on small
distances, and which contribution in general should be determined by
perturbation theory.
    Setting the energy value of general gluon field as   (this value
is unknown and therefore is the phenomenological parameter), we shall
obtain, that in the hadron rest-frame its mass equals:
\begin{equation}
         M_a = \epsilon_0 +\epsilon_1+\ldots+\epsilon_n
\end{equation}
\noindent
where $\epsilon_i, i=1,\ldots,n$ are defined as eigenvalues of Dirac
stationary equations (2) with potentials $U_i((\vec{q_i} - \vec{Q_i})_2)$and
quark current masses $\mu_i$ values.  In the quasi-independent generalized
quark fields $\Psi_i(Q,q_i )$ approximation we shall consider the simplest
    from the present hadron systems - two-particles mesons. All the more that
the produced numerical calculations of mass spectra in the framework of the
given model can be compared with a great number of very precise experimental
data [6].  Let us make use of the supposed spherical symmetry of scalar
potential and distinguish the stationary states of fermion components with
the definite values of total momentum$j_i$. Thus, the $i-$th quark/antiquark
state in the hadron will be determined by its spherical orbital $\Psi_{j_i}$.
The main difficulty in the present model consists in the reduction of a
number of a possible states with different values of $j_1$ and $j_2$.
     For two-particles mesons with different values of $J_{PC}$ ,built of
quark and antiquark with masses $\mu_1$ and $\mu_2$ the following selection
rules
have been found  [4] :
\begin{eqnarray}
&j_1 = j_2 = J+1/2\ \  &{\rm if}\ \   J=L+S  ;     \nonumber \\
&j_1 = j_2+1=J+3/2\ \  &{\rm if}\ \   J\neq L+S, \mu_1 \leq \mu_2
\end{eqnarray}
    The eigenenergy of the gluon field is also found to be dependent on
relative orbital momentum $L$ of quark and antiquark and their total spin $S$.
Dividing it in two parts between quark and antiquark, we can define the
mass spectral function (term) of quark and antiquark in the meson in the
following way:
\begin{equation}
         E_i (n_i ,j_i)=c[1+(-1)^{L+j_i-1/2}]+E'_i(n_i,j_i)
\end{equation}
\noindent
where $c=0.07$ GeV and the value of $E'_i(n_i, j_i)$ is found from the
approximate solution of squared Dirac equation   (2)   with the accont for
the phenomenological constants $L'=L$ and $\delta(J) \sim J$. For
two-particles mesons in the basic equation (2) with the scalar interaction
$S_i(r)$ one can include the vector interaction $V_i(R)$, where $r$  is the
relative distance between quark and antiquark, that is, make the substitution
in the equation (2):  \begin{equation} U_i(r) \rightarrow
\gamma_0V_i(r)+S_i(r) \end{equation} With the account for the selection rules
        (4), the mass  of the $n^{2S+1}L_J$ -meson state is calculated with the
help of the formula:  \begin{equation} M(n^{2S+1}L_J) = E_1 (n,J_1) + E_2(n,
   j_2), \end{equation}
\noindent
   where $n=1+n_r$, $n_r$ is the radial quantum number,
   moreover, the quantum numbers of quark and antiquark coincide.

\section { Light mesons mass spectra in the approximation       %3
                of quarks zero current masses and absence of
                    quasi-Coulomb interaction }
\noindent

    In some cases one can solve the equation (6) exactly and find
the analytical expression for quark and antiquark terms in mesons.
So, for example, it turns out well in the approximation of zero values of
quark current masses and absence of quasi-Coulomb interaction between them.
     Let us consider in this section the calculation of mass spectra of
$n^{2S+1}L_j$ -mesons, built of $u-, d-$ quarks and antiquarks. In this case
one
should expect, that a good approximation will be the following one:
$\mu_1 \simeq \mu_2 \simeq 0$ , where $\mu_1, \mu_2$ - current values of quark
and
antiquark masses. Let us also take into account, that one should keep  the
quasi-Coulomb perturbative term $-4\alpha_s /3r$ only at small distances,
therefore we can reject it for comparatively large dimensions of bound
states.  In this case the equality $L'\simeq L$ takes  place. Then it is not
difficult to obtain from equation (6) for quark and antiquark spectral
function the folowing formula:
\begin{equation} E(n,j)=c[1+(-1)^{L+j-1/2}]+
         \sqrt {\sigma(2n+L+j-1/2)}
         \end{equation}
    As far as we know, the formula (8) is the best one according to its
precision and fitting
parameters number for the light mesons and reproduces, together with the
selection rules (4), the the whole spectrum of states built of $u-, d-$
quarks and antiquarks [5]. The relative error of calculations has a range
of$\sim$ 1\% practically for all mesons, and only in some cases it reaches
3-4\%, that can be recognized as a good agreement, taking into account the
approximate character of formula (8). For the radial excitations of mesons
this formula in some approximation corresponds to the Veneziano-Namby
formula [6-8]. As can be seen from formula (8) the simple linear dependence
    of Chew-Frautschi [9] for Regge trajectories is valid only for some meson
states, for example for vector $1^{--}$ -mesons and their orbital
excitations.  In this case $\alpha'_R =1/8\sigma$. The same connection
between $\alpha'_R$ and $\sigma$ from other considerations was obtained in
the work [10]. Let us remind, that in the model of Nambu-Goto strings for the
hadrons $\alpha'_R$ and $\sigma$ are connected by the relation:
$\alpha'_R=1/2\pi\sigma$

\section{ Calculation of the parameters for the interaction  %4
              potential between quark and antiquark }

\noindent

    In the general case it is necessary to make use of the numerical
methods for the solution if the basic equation of the model. The main aim
in the present work is the determination of the parameters of the interaction
potential between quark and antiquark for all the known flavours,taking
into account quasi-Coulomb interaction. Without taking into consideration
quasi-Coulomb interaction, as it was noted in ref.[ 4 ],there had been
observed a systematic decrease of the slope parameter of the linearly rising
potential, when one proceeds from light to heavy quarks. This could be
connected  both with the neglect of the contribution of the quasi-Coulomb
interaction, and with an explicit dependence of the slope parameter of the
linearly rising potential on quark and antiquark flavour. With the help of
the numerical method of the solution of the problem on the eigenvalues of
the basic equation and following calculation of the masses of both light
and heavy vector mesons, we have found, that in the range of 1-2\% of the
relative error, which in most cases does not exeeds the experimental
one, with which  mesons masses spectra are determined,the hypothesis
of the univeresality for the slope parameter $\sigma$ of the scalar linearly
rising parameter takes place. The value of $\sigma$ , have been obtained, is
found to be 0.21 GeV$^2$, that corresponds to 1.06 GeV/Fm. The $\alpha_s$
value, which enters as a multiplier in the quasi-Coulomb term was $\sim$0.35
and weekly decreases when one proceeds from light to heavy quarks. This value
is consistent with the the generally accepted $\alpha_s$ values in
sufficiently broad domain of the distances  between interacting quark and
antiquark, however the further calculations are necessary for the
confirmation of the logarithmic dependence of $\alpha_s$ on the distance. The
quarks current masses thus also correspond to the values, found earlier in
the majority of works:  $m_{u,d}$ = 8 MeV, $m_s$ =1380 MeV, $m_b$ =4730 MeV.

\section{              Conclusion }
\noindent

    In the framework of the model with the phenomenological selection rules,
proposed in work [4] the parameters of the interaction potential between
quark and antiquark for all known up to now quark flavours are found.
It is found, that in the range of experimental errors, with which the mesons
masses are obtained, the universality hypothesis for the slope parameter of
the linearly rising potential, proposed earliar in works [11-13], takes
place. The $\sigma$ value is found to be 0,21 GeV$^2$. The increase of the
interaction constant on small distances $\alpha_s$ when one preceeds from
heavy to light quarks have been observed. The obtained results confirm the
approximate analytic formulae (8) for the mesons masses spectra, found in the
works [4] and at the same time it show, that the obtained on the base of
these formulae the parameter $\sigma$ values are have been diminished ones
and effectively take into account the contribution of the quasi-Coulomb
interaction.  The arthors are gratefull to V.I.Savrin and A.M.Snigirev for
usefull discussions.

%\section{                References}

\end{document}